\documentclass[10pt,conference]{IEEEtran}
\IEEEoverridecommandlockouts
\usepackage{cite}
\usepackage{amsmath,amssymb,amsfonts}
\usepackage{algorithmic}
\usepackage{graphicx}
\usepackage{textcomp}
\usepackage{xcolor}

\usepackage{hyperref}
\usepackage{booktabs}
\usepackage{caption}
\usepackage{subcaption}

\newcommand{\bn}[1]{\textsf{\textbf{#1}}}
\newcommand{\NA}{\multicolumn{1}{c}{---}}

\def\BibTeX{{\rm B\kern-.05em{\sc i\kern-.025em b}\kern-.08em
    T\kern-.1667em\lower.7ex\hbox{E}\kern-.125emX}}
\begin{document}

\title{Refactoring Programs Using Large Language Models with Few-Shot Examples
\thanks{This work was supported by the Japan Society for the Promotion of Science (JSPS) KAKENHI Grant Number JP23H03508.}
}

\author{
\IEEEauthorblockN{Atsushi Shirafuji\IEEEauthorrefmark{2}, Yusuke Oda\IEEEauthorrefmark{3}, Jun Suzuki\IEEEauthorrefmark{3}, Makoto Morishita\IEEEauthorrefmark{4}, Yutaka Watanobe\IEEEauthorrefmark{2}}
\IEEEauthorblockA{
\IEEEauthorrefmark{2}\textit{University of Aizu, Japan}\\
\IEEEauthorrefmark{3}\textit{Tohoku University, Japan}\\
\IEEEauthorrefmark{4}\textit{NTT Communication Science Laboratories, Japan}\\
Email: \{m5261161, yutaka\}@u-aizu.ac.jp, \{yusuke.oda.c1, jun.suzuki\}@tohoku.ac.jp, makoto.morishita@ntt.com
}
}

\maketitle

\begin{abstract}
A less complex and more straightforward program is a crucial factor that enhances its maintainability and makes writing secure and bug-free programs easier.
However, due to its heavy workload and the risks of breaking the working programs, programmers are reluctant to do code refactoring, and thus, it also causes the loss of potential learning experiences. 

To mitigate this, we demonstrate the application of using a large language model (LLM), GPT-3.5, to suggest less complex versions of the user-written Python program, aiming to encourage users to learn how to write better programs.
We propose a method to leverage the prompting with few-shot examples of the LLM by selecting the best-suited code refactoring examples for each target programming problem based on the prior evaluation of prompting with the one-shot example.

The quantitative evaluation shows that 95.68\% of programs can be refactored by generating 10 candidates each, resulting in a 17.35\% reduction in the average cyclomatic complexity and a 25.84\% decrease in the average number of lines after filtering only generated programs that are semantically correct.
Furthermore, the qualitative evaluation shows outstanding capability in code formatting, while unnecessary behaviors such as deleting or translating comments are also observed.
\end{abstract}

\begin{IEEEkeywords}
code refactoring, large language models, few-shot prompting, software complexity, programming education
\end{IEEEkeywords}

\section{Introduction}
\label{sec:introduction}

\begin{figure}[h]
    \centering
    \includegraphics[width=\linewidth]{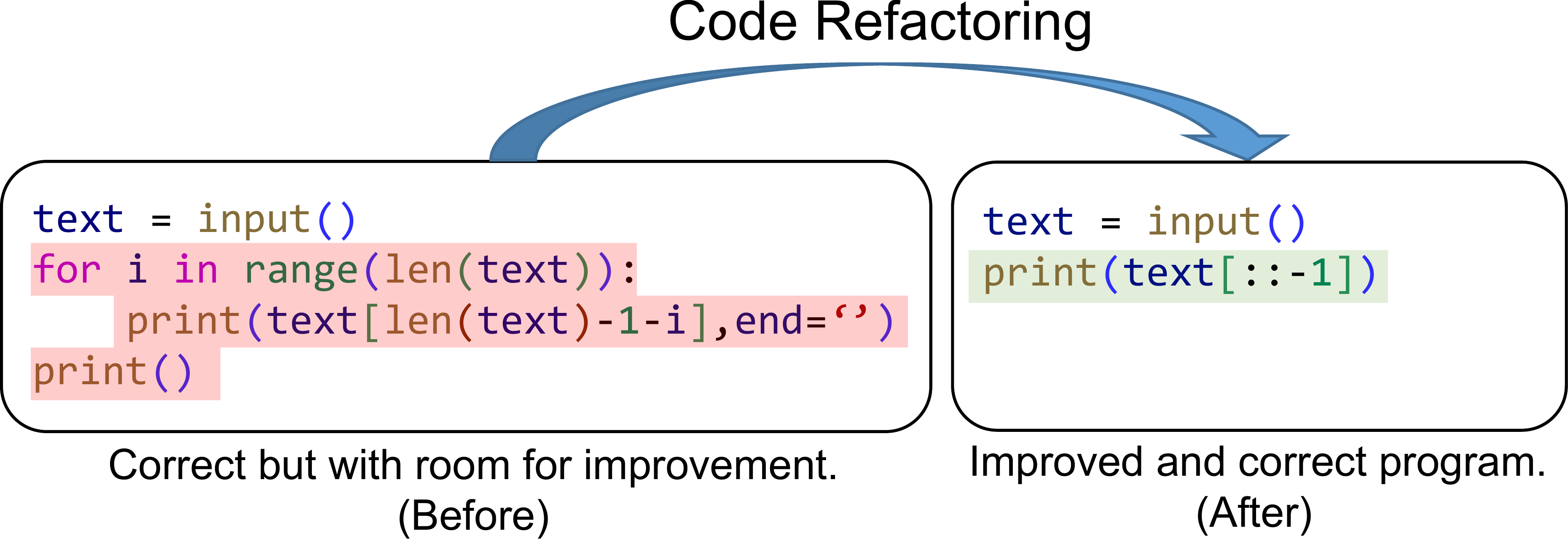}
    \caption{Example of code refactoring to improve a correct program with room for improvement.}
    \label{figure:code-refactoring}
\end{figure}

Programmers and learners often write unreadable, redundant, or complicated programs because they lack the knowledge to write less complex programs or are in a hurry to meet minimum requirements by sacrificing the complexity.
Writing readable and maintainable programs from the beginning is difficult for not only novices but also experts.
They often iteratively make minor modifications to the previous version of the program to improve the code readability and maintainability, called \textit{code refactoring} (Figure~\ref{figure:code-refactoring}).

Complicated and unreadable programs have risks of making them harder to maintain (i.e., difficult to add, modify, or remove functions) and to hand over to a successor because the program can be understood by the author only, as well as the risks of making bugs and the difficulty of finding and fixing the potential bugs.

Programmers sometimes have no choice but to write such complicated and unreadable programs because they lack the knowledge to write the program in a less complex and more straightforward way than the one that comes to their mind, or they intentionally leave them complicated and unreadable because of the high cost of modifying them (e.g., lack of time for delivery) although they recognize that they should deal with them.
According to one of the definitions of code refactoring by Fowler~et~al.~\cite{fowler1999refactoring}, ``the process of changing a software system in such a way that it does not alter the external behavior of the code yet improves its internal structure,'' it should keep the external behavior of the previous version of the program.
In fact, code refactoring also has a risk of breaking the already working program, which is another reason programmers are reluctant to do code refactoring.

In recent years, large language models (LLMs) trained on texts in programming languages as well as natural languages~\cite{chen2021codex,chowdhery2022palm,christopoulou2022pangucoder,fried2022incoder,li2022alphacode,nijkamp2022codegen} have shown the potential to support programming in both programming education~\cite{finnie2022robots,sarsa2022exercise,wu2021prototransformer,zhang2022repair} and software development~\cite{svyatkovskiy2020intellicode,vaithilingam2022copilot,xu2022in-ide}.
From a prompt engineering perspective, White~et~al.~\cite{white2023prompt} proposed several prompt patterns for code refactoring to support software engineering activities.
As the most related work, Madaan~et~al.~\cite{madaan2023improving} used two popular LLMs, Codex and CodeGen, to improve the program's time efficiency.
In another work, Madaan~et~al.~\cite{madaan2023code-readability} used Codex to improve the readability of variable names and comments.
However, to the best of our knowledge, no prior work has demonstrated code refactoring to reduce the software complexity using LLMs.

To address the issue rising in both the fields of software development and programming education, we propose the use of an existing LLM, GPT-3.5, to suggest a complexity-improved version of a user-written complex Python program to motivate them to learn how to write better programs, as well as supporting programming instructors by reducing their workloads to think and answer the questions for each learner.
To leverage the prompting with few-shot examples\footnote{In this paper, we denote \textit{n-shot prompting} for prompting with n-shot examples.} of the LLM, we propose a method to select the best-suited code refactoring examples used for few-shot prompts in each target programming problem based on the prior evaluation results of one-shot prompting for each code refactoring example.
Since the LLM has the potential to break the input program, the generated programs are validated, and only functionally correct programs are suggested to a user.

In the experiments, we collect functionally correct Python programs to be refactored from a set of 44 introductory programming problems provided on Aizu Online Judge (AOJ)~\cite{watanobe2004aoj} and randomly select 20 unique programs in each programming problem.
We generate 10 code refactoring candidates for each of the correct programs and verify if the generated program is syntactically and semantically correct.
We demonstrate the applicability of the LLM to generate less complex programs aligned to the user-written program, evaluated both quantitatively and qualitatively.

The contributions of this work are as follows:
\begin{itemize}
    \item We demonstrate that the LLM can generate correct and less complex programs for a majority of the input programs.
    \item We propose leveraging few-shot prompting by selecting refactoring examples to help guide the LLM to better align with user-written programs. 
    \item Our quantitative and qualitative evaluations exhibit the LLM's performance in code refactoring and its potential to assist programming.
    \item We discuss the limitations of the current approach, providing insights for future research in leveraging LLMs for code refactoring.
\end{itemize}

\section{Related Work}
\label{sec:related-work}

\begin{figure*}
    \centering
    \includegraphics[width=\linewidth]{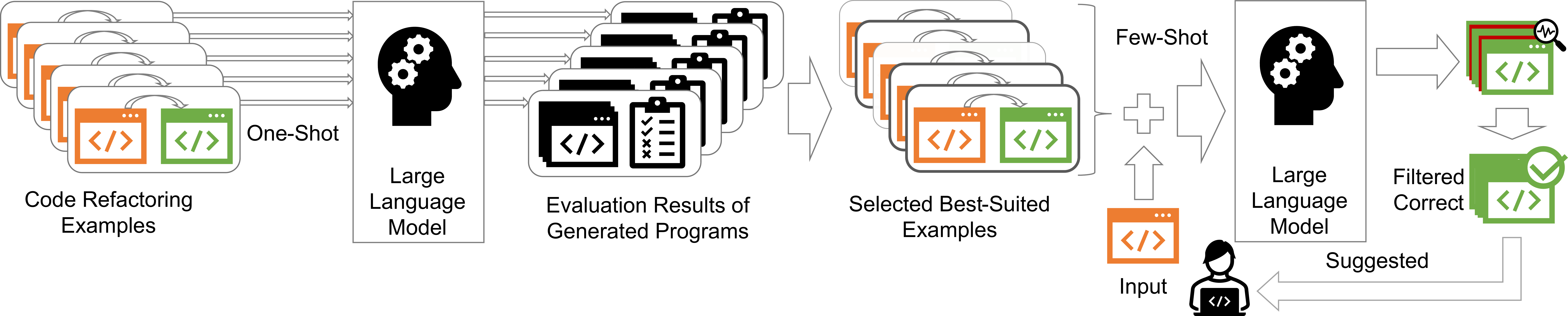}
    \caption{Illustration of the proposed approach selecting the best-suited code refactoring examples for few-shot prompting for each programming problem based on the performance of one-shot prompting. Only the filtered programs are suggested to a user.}
    \label{figure:methodology}
\end{figure*}

Code refactoring is a well-known practice in software development that aims to improve the internal structure and readability of a program without changing its external behavior~\cite{fowler1999refactoring}.
Most code refactoring tasks are performed manually by developers based on experience and best practices~\cite{murphyhill2012refactor}.
However, there have been several research efforts to assist or automate code refactoring using various techniques.

One important target of code refactoring is code clone or code duplication, which refers to reusing several portions of code across different parts of the codebase instead of defining modular functions.
Code clones increase complexity, reduce maintainability, and violate the \textit{Don't Repeat Yourself} (DRY) principle.
Refactoring the duplicated or near-duplicated code fragments makes the program more concise and easier to maintain~\cite{tsantalis2015, tsantalis2017}, and code clone detection techniques~\cite{kamiya2002ccfinder} can also be applied.

Several different approaches have been proposed to assist code refactoring.
WitchDoctor~\cite{foster2012witchdoctor} is a tool that can detect refactoring behaviors of programmers and automatically complete the refactoring while they are being performed.
Blue-Pencil~\cite{miltner2019bluepencil} empowers the refactoring feature of Visual Studio IntelliCode\footnote{\url{https://devblogs.microsoft.com/visualstudio/refactoring-made-easy-with-intellicode/}.} to suggest repetitive edits automatically.
Search-based techniques, which find code fragments that can improve program readability or complexity, have also been applied in refactoring suggestions~\cite{seng2006, harman2007, adler2021}.

In addition to automated code refactoring approaches mentioned above, refactoring is also used in a more narrow context within integrated development environments (IDEs).
In IDEs, refactoring often refers to the process of automatically renaming identifiers, such as classes, functions, and variables, throughout the codebase.
Many IDEs, such as Visual Studio\footnote{\url{https://visualstudio.microsoft.com/}.}, Eclipse\footnote{\url{https://www.eclipse.org/}.}, and IntelliJ~IDEA\footnote{\url{https://www.jetbrains.com/idea/}.}, provide built-in functionality to facilitate this renaming process, ensuring consistent updates of identifiers throughout the codebase.

It is worth noting that code refactoring can be considered a code-to-code generation task in the natural language processing field since a potentially complex program written in a programming language is converted into a less complex program written in the same language.
While code refactoring aims to improve already correct programs, code repair focuses on converting incorrect programs into correct ones~\cite{rahman2021repair, matsumoto2021repair, shirafuji2023repair}.
LLMs, such as Codex~\cite{chen2021codex}, have shown the capability of repairing programs~\cite{joshi2022repair,pearce2021securitybugs,prenner2021repair,zhang2022repair}.
Code editing~\cite{chakraborty2022codit,li2022code-editor,zhang2022codit-t5} is a more generalized task that involves learning the code editing behavior.
Pre-trained models can be fine-tuned on downstream tasks such as code repair~\cite{li2022code-editor,zhang2022codit-t5}. 

In recent work, White~et~al.~\cite{white2023prompt} proposed a catalog of prompt patterns for software engineering activities using ChatGPT as a representative LLM. In particular, they provided six prompt patterns for code refactoring tasks, such as making a given program follow certain coding principles.
As the most related work, Madaan~et~al.~\cite{madaan2023improving} demonstrated the capability of LLMs (i.e., Codex~\cite{chen2021codex} and CodeGen~\cite{nijkamp2022codegen}) to improve the time efficiency of programs by proposing a large-scale dataset consisting of (slower, faster) pairs of programs written by the same user.
Similarly, Madaan~et~al.~\cite{madaan2023code-readability} used Codex to improve the readability of variable names and comments by leveraging the iterative refinement of the generated programs using the feedback generated by Codex, named \texttt{Self-Refine}.
These works are closely related to ours because they focus on improving already correct programs in terms of performance or readability by keeping the functional correctness.
In contrast, our work focuses on reducing program complexity.

\section{Methodology}
\label{sec:method}

\subsection{Overview}
\label{sec:method:overview}

The proposed methodology is illustrated in Figure~\ref{figure:methodology}.

Firstly, we define code refactoring examples used in prompts.
Secondly, we evaluate each example in each programming problem, and based on the performance of each example, we select the best-suited examples for each programming problem.
Thirdly, the user-written program is passed to the LLM with the selected few-shot examples, and the LLM generates several refactoring candidates.
Finally, we validate the generated programs using an automatic judge system, and only functionally correct programs are suggested to the user.

\subsection{Examples Preparation}
\label{sec:method:preparation}

\textit{Few-shot prompting}~\cite{brown2020gpt3} provides a few examples for LLMs to demonstrate the expected inputs and outputs of conversations, whereas zero-shot prompting provides no examples.

\begin{figure}[h]
    \centering
    \includegraphics[width=\linewidth]{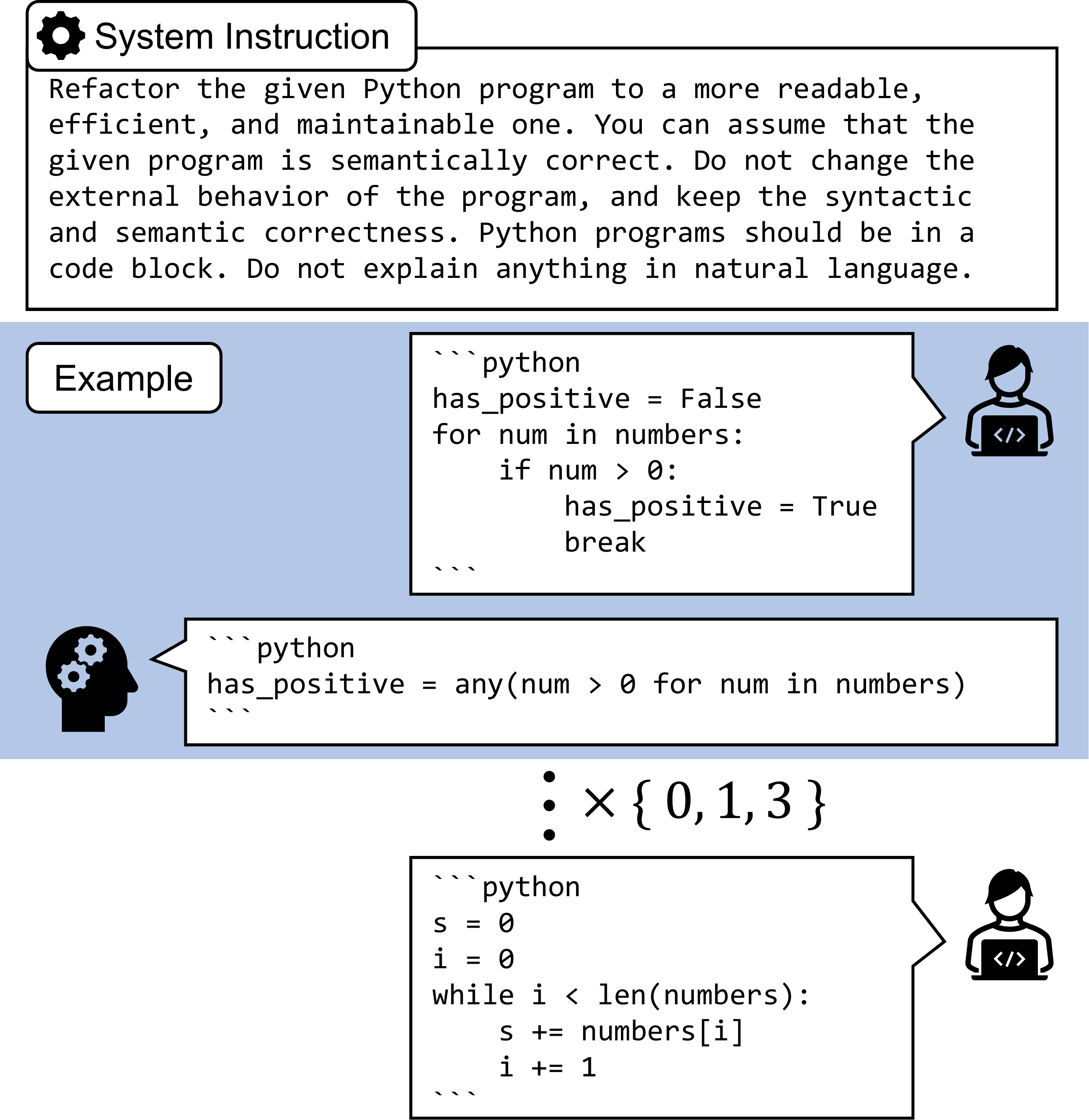}
    \caption{Illustration of prompting consisting of (1) a system instruction, (2) zero/one/few-shot examples, and (3) the user's input program. The conversation in blue is the code refactoring example.}
    \label{figure:few-shot}
\end{figure}

Figure~\ref{figure:few-shot} illustrates the prompting in this work. 
The system instruction and the user's program are always passed to the LLM.
No examples are passed in the zero-shot, one in the one-shot, and three in the few-shot (i.e., 3-shot).
The program on the LLM (left) side is not actually generated by the LLM but prepared by us manually.
It imitates that it is generated to demonstrate the generation of the LLM.

We manually define the following 10 code refactoring examples, aiming to reduce the software complexity by utilizing the defined functions and statements in Python, as well as some techniques that make the program more readable but do not directly reduce the software complexity.

\begin{itemize}
    \item Use a formatted string.
    \item Use a built-in (i.e., radians) function.
    \item Use a logical operator instead of a nested if.
    \item Use a for-loop instead of a while-loop.
    \item Use list comprehension instead of a for-loop.
    \item Use the map function instead of list comprehension.
    \item Use a throwaway variable.
    \item Use the enumerate function instead of the range function.
    \item Use the zip function instead of the range function.
    \item Use a ternary operator instead of an if-branch.
\end{itemize}

\subsection{Examples Evaluation}
\label{sec:method:evaluation}

In this phase, each code refactoring example is used for one-shot prompting and evaluated by a validator.

Let $\mathbf{E} = \{ (\text{original}, \text{refactored}) \}$ be a set of code refactoring examples, where $(\text{original}, \text{refactored}) \in \mathbf{E}$ is a pair consisting of a correct program with room for improvement and its refactored version.
Hereafter, let us denote $e = (\text{original}, \text{refactored})$ for short, and thus, $e \in \mathbf{E}$. 
Note that the number of code refactoring examples in $\mathbf{E}$ is a hyper-parameter.
We use $|\mathbf{E}| = 10$ in our experiments, as defined in Section~\ref{sec:method:preparation}.

Let $\mathbf{P}$ be a set of programming problems.
Then, each programming problem $P \in \mathbf{P}$ has a set of original correct programs written by users $\mathbf{X}_P = \{ x \}$.
Moreover, $\mathbf{Y}_{ex} = \{ y \}$ is a set of code refactoring candidates generated by a pre-trained LLM using a code refactoring example $e \in \mathbf{E}$ and an input program $x \in \mathbf{X}_P$.
In our experiments, we use $|\mathbf{X}_P| = 20$ programs for $|\mathbf{P}| = 44$ programming problems and generate $|\mathbf{Y}_{ex}| = 10$ code refactoring candidates. 

A validator $\mathcal{V}_P$ to validate the correctness of a generated program $y$ for the programming problem $P$ can be denoted as Formula~\ref{formula:validator}.
\begin{equation}
    \mathcal{V}_P(y) = 
        \left\{
        \begin{array}{ll}
        1 & \text{if $y$ solves $P$,} \\
        0 & \text{otherwise.}
        \end{array}
        \right.
    \label{formula:validator}
\end{equation}
In practice, the validator actually executes the given program using test cases and returns the correctness in binary, as described in Section~\ref{sec:experiment:environment}.

Then, for each code refactoring example $e \in \mathbf{E}$ and each programming problem $P \in \mathbf{P}$, the score $S_{eP}$ can be denoted as Formula~\ref{formula:problem_score}.
This score indicates how suitable the code refactoring example $e$ is for the programming problem $P$.
\begin{equation}
    S_{eP} = \sum_{x \in \mathbf{X}_P} \sum_{y \in \mathbf{Y}_{ex}} \mathcal{V}_P(y)
    \label{formula:problem_score}
\end{equation}

Finally, for each code refactoring example $e \in \mathbf{E}$, the cumulative score $S_e$ can be denoted as Formula~\ref{formula:example_score}.
This score reflects how suitable the code refactoring example $e$ is overall.
\begin{equation}
    S_e = \sum_{P \in \mathbf{P}} S_{eP}
    \label{formula:example_score}
\end{equation}

\subsection{Examples Selection}
\label{sec:method:selection}

Let $\text{argtopk} (\mathbf{x}, k)$ a function to return the arguments at which the function returns the top-$k$ values, the best-performed $k$ code refactoring examples $\mathbf{E}_P$ for each programming problem $P \in \mathbf{P}$ is calculated by Formula~\ref{formula:examples}.
\begin{equation}
    \mathbf{E}_P = \text{argtopk}_e (1000 S_{eP} + S_e, k)
    \label{formula:examples}
\end{equation}

After selecting the best-performed code refactoring examples for few-shot prompting, we construct the few-shot prompts for the LLM.
As a few-shot prompting, we adopt 3-shot prompting, which provides $k = 3$ sets of best-performed code refactoring examples.

\section{Experimental Setup}
\label{sec:experiment}

We demonstrate the applicability of an LLM in code refactoring to reduce software complexity.
In this section, we describe the original dataset consisting of target programs to be refactored, the model from the representatives of off-the-shelf LLMs used to generate the refactored programs, and the objective metrics to evaluate the effectiveness.

\subsection{Dataset}
\label{sec:experiment:dataset}

We construct a dataset consisting of the original programs to be refactored by an LLM.

\subsubsection{Data Collection}
\label{sec:experiment:dataset:collection}

To construct a dataset, we collect a variety of correct programs written in Python3 from AOJ, an online judge system where users submit programs to solve given programming problems.
AOJ provides approximately 3,000 programming problems and stores 8,000,000 programs submitted by 100,000 users, including wrong or incomplete programs~\cite{watanobe2022aoj}.
The source code submitted to AOJ is available for research or educational purposes and can be downloaded from the official source archive\footnote{\url{http://developers.u-aizu.ac.jp/index}.} or via public datasets such as CodeNet~\cite{puri2021codenet} or CodeContests~\cite{li2022alphacode}.

We limit the target problems from a popular course problem named \textit{Introduction to Programming I} (ITP1)\footnote{\url{https://onlinejudge.u-aizu.ac.jp/courses/lesson/2/ITP1/all}.}, which has 44 introductory programming problems, ranging from basic operations of input/output to class definitions.

Each submission of a program on AOJ has a verdict resulting from executing hidden test cases by the judge system, such as \textit{Accepted}, \textit{Wrong Answer}, or \textit{Runtime Error}.
We only use the submissions judged as \textit{Accepted} because we only focus on the code refactoring task to improve correct but complex programs.

By the above collecting conditions, we collect 296,885 correct Python programs from 44 programming problems on AOJ.

\subsubsection{Preprocessing}
\label{sec:experiment:dataset:preprocess}

\begin{table}[h]
\begin{center}
\begin{minipage}{\linewidth}
    \caption{Statistics of collected programs in each of the data preprocessing phases. LOC indicates the average number of lines, Chars and Tokens indicate the average number of characters and tokens, respectively, and CC indicates the average cyclomatic complexity.}
    \label{table:data-statistics}
    \begin{tabular*}{\textwidth}{@{\extracolsep{\fill}}lrrrrr@{\extracolsep{\fill}}}
        \toprule
        & \#Programs & LOC & Chars & Tokens & CC \\
        \midrule
        Collected & 296,885 & 14.59 & 332.56 & 108.79 & 5.85 \\
        Unique & 161,670 & 15.24 & 346.19 & 111.88 & 6.01 \\
        Filtered & 158,081 & 14.60 & 329.40 & 107.26 & 5.79 \\
        Selected & 880 & 14.82 & 331.35 & 106.15 & 5.65 \\
        \bottomrule
    \end{tabular*}
\end{minipage}
\end{center}
\end{table}

After collecting the correct programs from AOJ, we preprocess the data to select partial programs for the experiments in the following phases.
The statistics of the data in each phase are summarized in Table~\ref{table:data-statistics}.
Also, the metrics used in the table are explained in Section~\ref{sec:experiment:metrics}.

\paragraph{Duplicate Deletion}
Firstly, we remove duplicate programs in each problem.
The duplicate detection is based on the characters of the raw source code, neither based on tokens nor trees.
Therefore, the programs are considered different if only one space or empty line is different.
After removing the duplicated programs, the number of unique programs is 161,670.

\paragraph{Outlier Deletion}
Secondly, we remove outliers in each problem.
The outlier criterion is $2\sigma + \mu$, where $\sigma$ is the standard deviation, and $\mu$ is the mean of the number of tokens in programs in each problem.
After removing outlier programs, the number of filtered programs is 158,081.

\paragraph{Randomly Selection}
Finally, we randomly select 20 programs for each problem, and the final dataset contains 880 programs in total (20 programs $\times$ 44 problems $=$ 880 programs).

\subsection{Model}
\label{sec:experiment:model}

To demonstrate the few-shot prompting using the pre-trained LLM, we use the representative LLM from the off-the-shelf models.
We use the \texttt{gpt-3.5-turbo} engine served from the OpenAI API\footnote{\url{https://platform.openai.com/docs/models/gpt-3-5}.}, often referred to as ChatGPT or GPT-3.5 models.
More precisely, in this work, we use \texttt{gpt-3.5-turbo-0301}, a snapshot version from March 1st, 2023, for reproducibility.
\texttt{gpt-3.5-turbo} is an InstructGPT~\cite{ouyang2022instruct-gpt}-based model, trained to follow user instructions and provide detailed responses using a technique called \textit{reinforcement learning from human feedback} to align with the user's instructions.
InstructGPT is also based on Codex~\cite{chen2021codex}, which is trained on massive source code.
Therefore, \texttt{gpt-3.5-turbo} has a high capability in understanding and generating programming languages as well as natural languages.

We set the \textit{temperature}, which determines the creativity of generated texts, to 0.2 to be slightly creative for all generations in this work.
We set the \textit{max\_tokens}, which limits the maximum number of tokens to generate, to 1,024.
Although \texttt{gpt-3.5-turbo} supports up to 4,097 tokens, it counts both the prompt and generation tokens.
Given that the average number of tokens in the input programs is 106.15, as shown in Table~\ref{table:data-statistics}, the limit of 1,024 tokens is sufficient for generating refactored programs.
We use the following system instruction to ask the LLM to do code refactoring:
\textit{``Refactor the given Python program to a more readable, efficient, and maintainable one. You can assume that the given program is semantically correct. Do not change the external behavior of the program, and keep the syntactic and semantic correctness. Python programs should be in a code block. Do not explain anything in natural language.''}
The system instruction is used in all generations, including zero-shot prompting.
In addition, as instructed in the system instruction, each Python program communicated with the LLM is represented as a code block, enclosed with three backticks.

\subsection{Metrics}
\label{sec:experiment:metrics}

Since our purpose is to suggest less complex programs as well as functionally correct programs for the user, we evaluate the generated programs from the perspective of correctness and complexity.
We also report other software metrics for analysis, such as Levenshtein distance (or edit distance), lines of code (LOC), the number of characters, the number of tokens, and the character-based similarity.

\paragraph{Compilability} is the syntactic correctness of the generated program, whether the program passes the compilation, denoted by $\text{Compilability} = P / (P + F)$, where $P$ is the number of programs that passed the compilation, and $F$ is the number of programs that failed in the compilation.
Since all the collected programs for code refactoring initially solved the problem, the compilability of the original programs is 100\%.

\paragraph{Pass@$k$} is used to evaluate the semantic (functional) correctness of the generated programs, proposed by Chen~et~al.~\cite{chen2021codex}.
Although the metric is designed to evaluate \textit{if the problem is solved} by generating $k$ samples, we adopt this metric to evaluate \textit{if the program is refactored} by generating $k$ samples.
Semantic correctness is validated by a virtual judge system, where hidden test cases are given to the program and considered correct if the program passes all the hidden test cases.
Pass@$k$ is denoted as Formula~\ref{formula:pass-at-k}, where $n \geq k$ is the number of samples and $c \leq n$ is the number of correct samples.
\begin{equation}
    \text{pass@$k$} := \mathop{\mathbb{E}}_{\text{Problems}} \left[ 1 - \frac{{\binom{n-c}{k}}} {\binom{n}{k}} \right]
    \label{formula:pass-at-k}
\end{equation}
We generate $n = 10$ samples for each program and report pass@$k$ at $k = 1, 10$.
In this work, pass@1 indicates the ratio of correct programs out of all generated programs.
Also, assuming to suggest the programs only validated as correct, pass@10 indicates the ratio of programs that at least one refactored program is suggested within the 10 generation tries.

\paragraph{McCabe's Cyclomatic Complexity (CC)} is a software metric to measure the program's complexity~\cite{mccabe1976complexity}.
CC increases with the number of if branches, for loops, and their nesting.
A smaller CC is preferable, and $\text{CC} < 10$ is categorized as a \textit{little risk} in software development.
We employ the radon\footnote{\url{https://github.com/rubik/radon}.} library to calculate the CC.

\paragraph{Chars} is a length based on the number of characters in the program.
This metric is highly influenced by the longer tokens, such as string literals, comments written in natural language, and white spaces, as well as the variable or function names.
However, this metric can reflect the raw length of the program compared to the other metrics.

\paragraph{Tokens} is a length based on the number of tokens in the program.
This metric can particularly reflect the number of elements (tokens) directly influencing the program's behavior, not affected by superficial texts that are already measured by the chars.
We utilize the official Python tokenizer\footnote{\url{https://docs.python.org/library/tokenize.html}.} to tokenize the programs.
In tokenization, we exclude comments and special tokens, such as newline, indent, and dedent, to better reflect the length of the semantic elements of the program.

\paragraph{Lines of Code (LOC)} is a length based on the number of lines in the program, including empty and comment lines.
Another metric named Source Lines of Code (SLOC) excludes empty and comment lines in counting.
Although either metric can be used in the experiments, we use LOC since it can reflect the number of insertions and deletions of empty and comment lines.

\paragraph{Distance} is a Levenshtein distance~\cite{levenshtein1966} calculated between the original and generated programs based on characters, also known as edit distance.
Each inserted, deleted, or substituted character is counted as one distance, whereas there is another method to count substitution as two distances (i.e., an insertion and a deletion).

\paragraph{Similarity} is a syntactic similarity between the original and generated programs based on characters using the Levenshtein distance.
In this work, the similarity of two programs, $a, b$, of length $|a|, |b|$, is defined as Formula~\ref{formula:similarity}, where $\text{Distance}(a, b)$ is the Levenshtein distance between $a$ and $b$.
\begin{equation}
    \text{Similarity}(a, b) = 1 - \frac{\text{Distance}(a, b)}{\max(|a|, |b|)}
    \label{formula:similarity}
\end{equation}

\subsection{Execution Environment}
\label{sec:experiment:environment}

To validate the semantic correctness of generated programs, we execute the generated program using test cases for each problem.
The program is judged correct if it passes all test cases and is incorrect otherwise, similar to unit testing in software development.
The test cases for each problem are available through AOJ API\footnote{\url{http://developers.u-aizu.ac.jp/index}.}.
However, several prior works mentioned that LLMs trained on public source code have a risk of generating malicious or vulnerable programs, which may harm the host computer~\cite{chen2021codex,pearce2021security}.
Therefore, we prepare a virtual judge system on our isolated sandbox environment to not be affected by the malicious generated programs.

\section{Results}
\label{sec:results}

\subsection{Quantitative}
\label{sec:results:quantitative}

\subsubsection{Main Results}

\begin{table}[h]
\begin{center}
\begin{minipage}{\linewidth}
    \caption{Pass@$k$, compilability, and CC of generated programs, along with the input programs. \textit{One-shot} indicates the best one-shot using the \textit{list comprehension} example that performed the best among one-shot prompting on both pass@1 and pass@10. \textit{Few-shot} indicates our proposed approach using 3-shot prompting. CC indicates cyclomatic complexity. The best score in each metric is in bold.}
    \label{table:pass_rates}
    \begin{tabular*}{\textwidth}{@{\extracolsep{\fill}}lrrrrrr@{\extracolsep{\fill}}}
        \toprule
        & Pass@1 & Pass@10 & Compilability & CC \\
        \midrule
        Input & \NA & \NA & 100.00\% & 5.65 \\
        \midrule
        Zero-shot & 87.44\% & 93.30\% & \bn{99.99\%} & 4.71 \\
        One-shot & 89.77\% & 94.77\% & 99.95\% & 4.78 \\
        Few-shot & \bn{91.11\%} & \bn{95.68\%} & 99.93\% & \bn{4.67} \\
        \bottomrule
    \end{tabular*}
\end{minipage}
\end{center}
\end{table}

Table~\ref{table:pass_rates} shows the main evaluation results of our proposed methodology.
Both pass@1 and pass@10 scores are improved by giving more examples in prompts, and the proposed 3-shot prompting resulted in the best performance.
In addition, the CC is improved (reduced) by 17.35\% compared to the input programs.
Although the compilability slightly decreases, we prioritize the higher scores of pass@$k$ because the semantically wrong programs cannot be suggested to the user even if they are compilable.

\subsubsection{Detailed Scores for Each Prompting}

\begin{figure}[h]
\centering
\includegraphics[width=\linewidth]{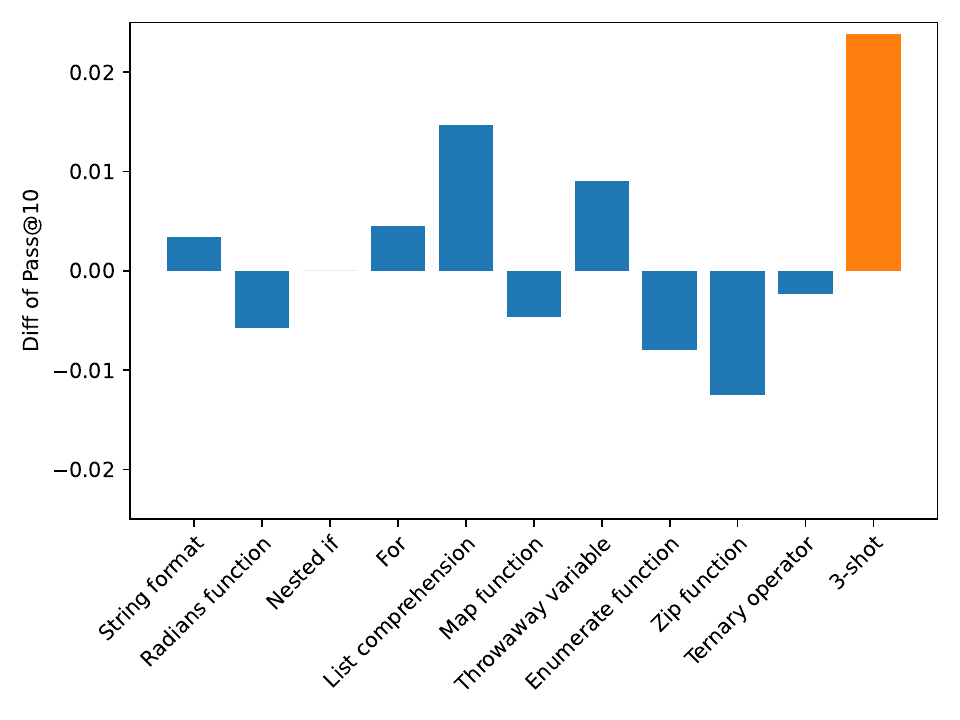}
\caption{The difference of pass@10 from zero-shot prompting for each prompting. Higher is better.}
\label{figure:pass10}
\end{figure}

\begin{figure}[h]
\centering
\includegraphics[width=\linewidth]{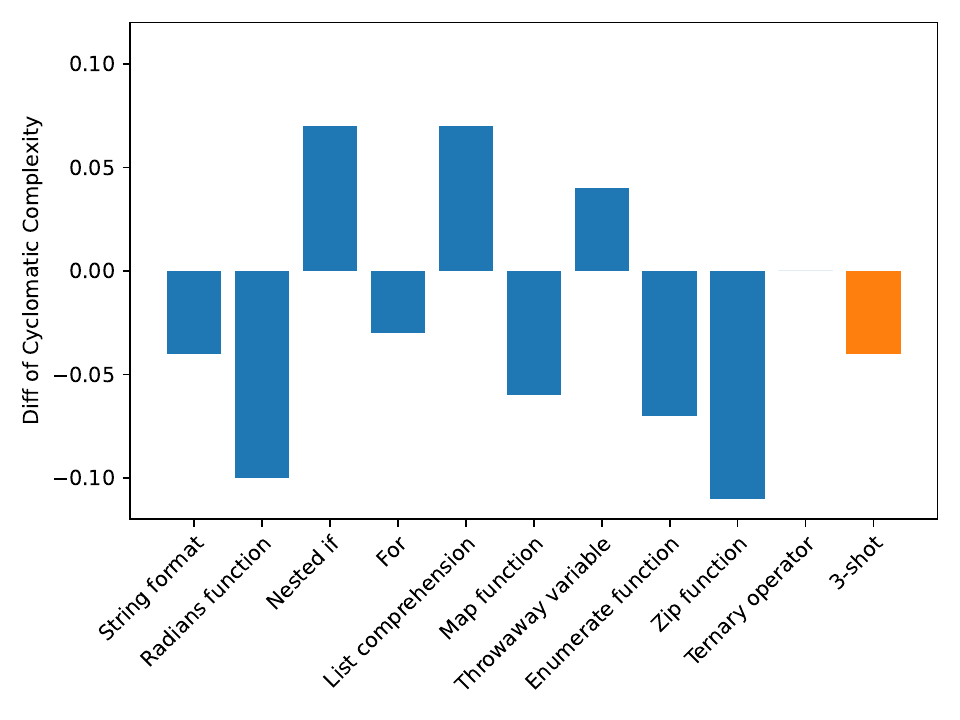}
\caption{The difference of CC from zero-shot prompting for each prompting. Lower is better.}
\label{figure:cc}
\end{figure}

Figure~\ref{figure:pass10} and Figure~\ref{figure:cc} show the detailed scores for each prompting, representing the difference from the zero-shot prompting.
While the 3-shot prompting yielded the best performance on pass@10, several one-shot prompting also performed better than the zero-shot prompting.
This variation of performance among the one-shot prompting highlights the need for selecting the best-suited examples for each problem.

\subsubsection{Conflict Between Pass@10 and Cyclomatic Complexity}

\begin{figure}[h]
\centering
\includegraphics[width=\linewidth]{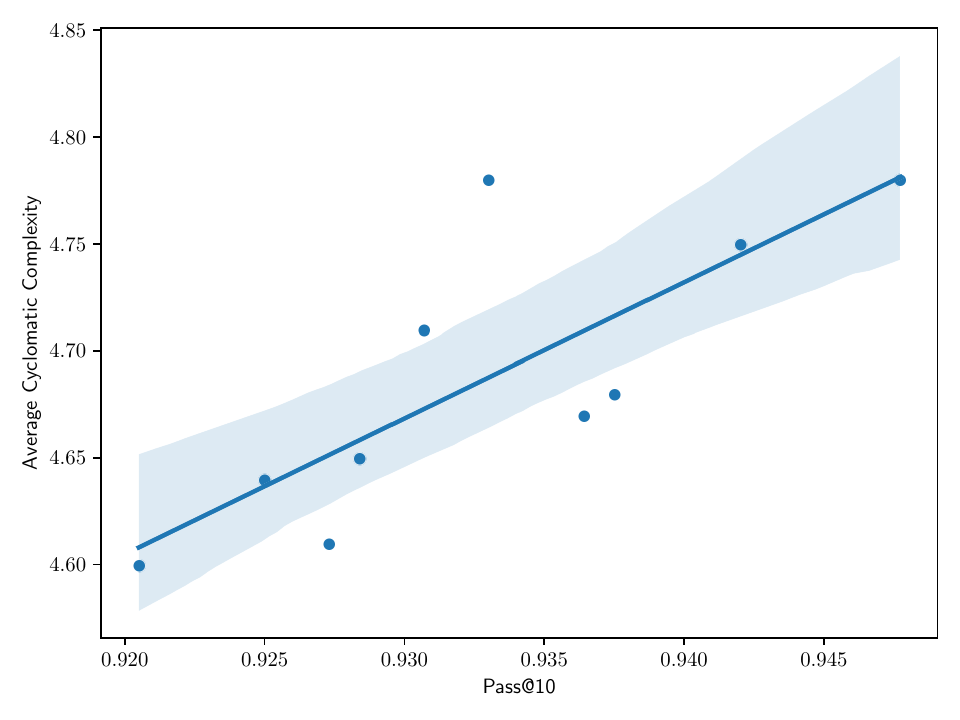}
\caption{A strong positive correlation (coefficient is 0.7916) between pass@10 and CC, indicating the difficulty of reducing the CC while keeping the pass@10 simultaneously.}
\label{figure:pass10-cc}
\end{figure}

When we refer to the relationship between pass@10 and CC, we observe that these two metrics conflict.
For Figure~\ref{figure:pass10}, the best (highest) one-shot prompting uses the \textit{list comprehension} example, and the worst (lowest) prompting uses the \textit{zip function} example.
In contrast, as shown in Figure~\ref{figure:cc}, the \textit{zip function} example that performed the worst in pass@10 is the best (lowest) in CC.
Similarly, the \textit{list comprehension} example that performed the best in pass@10 is the worst (largest) in CC.
This conflict of metrics is verified in Figure~\ref{figure:pass10-cc}.
We identify a strong positive correlation between the pass@10 and CC, as the correlation coefficient is 0.7916.
It indicates the difficulty of reducing the CC while keeping the pass@10 simultaneously.
However, to raise the availability of suggesting at least one candidate of code refactoring for each program, we prioritize raising the pass@10 in this work, and reducing the CC is our secondary importance.

\subsubsection{Decrease in Cyclomatic Complexity}

\begin{figure}[h]
\centering
\includegraphics[width=\linewidth]{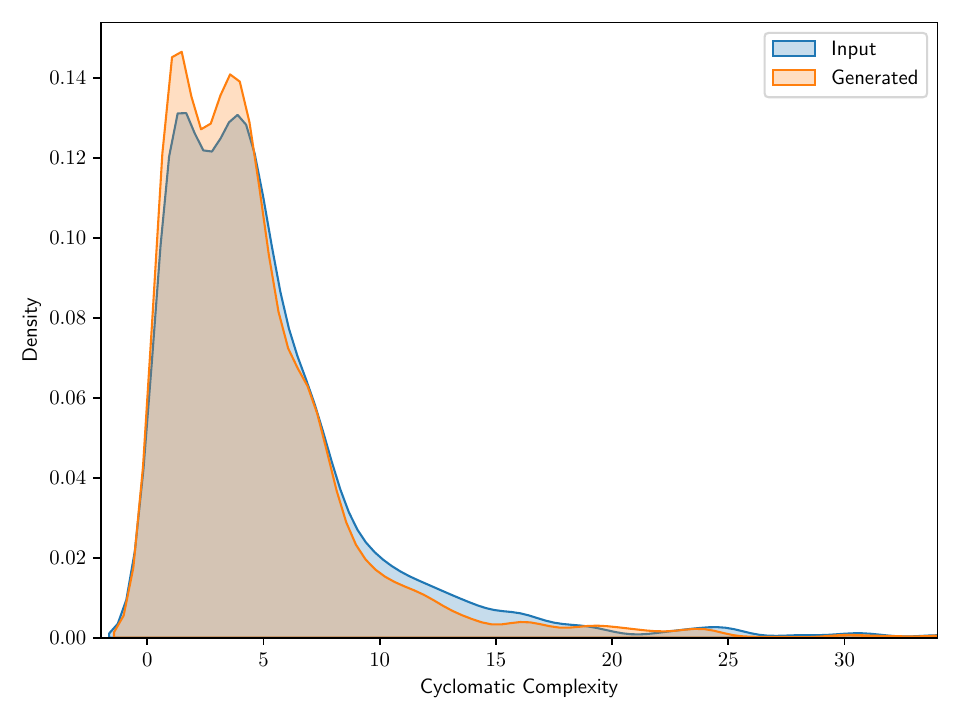}
\caption{Distributions on the CC for input and generated programs.}
\label{figure:cc-distribution}
\end{figure}

We test the difference in CC between the input and generated programs using the Wilcoxon signed-rank test, where the significance level is set to 0.05.
We find a statistically significant difference, as the P-value is less than 0.001.
As shown in Figure~\ref{figure:cc-distribution}, the CC in the generated programs generally decreases.

\subsubsection{Correlations in Levenshtein Distance}

\begin{figure}[h]
\centering
\includegraphics[width=\linewidth]{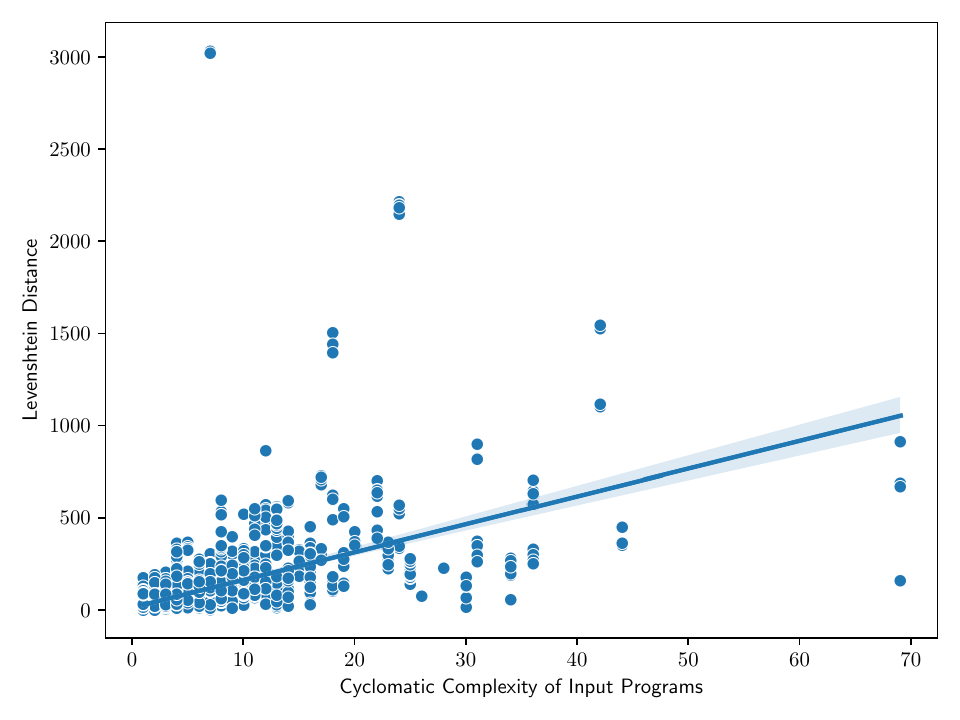}
\caption{A positive correlation (coefficient is 0.4886) between the CC of input programs and the Levenshtein distance, indicating that the LLM made larger modifications to the programs with higher complexity.}
\label{figure:cc-distance}
\end{figure}

\begin{figure}[h]
\centering
\includegraphics[width=\linewidth]{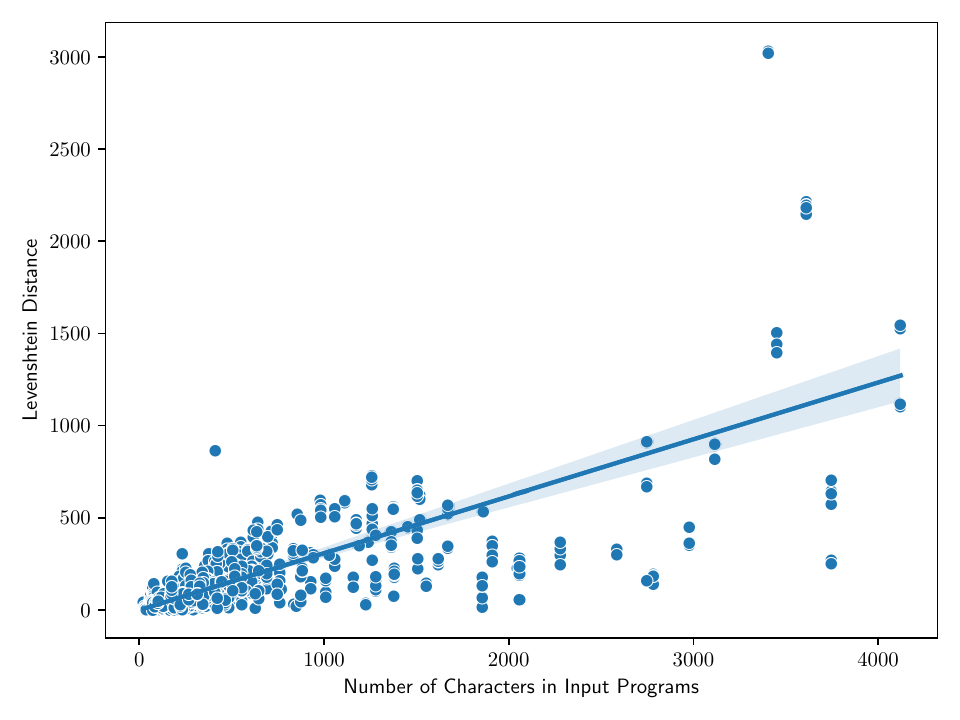}
\caption{A strong positive correlation (coefficient is 0.7442) between the number of characters in input programs and the Levenshtein distance, indicating that the LLM made larger modifications to longer programs.}
\label{figure:chars-distance}
\end{figure}

Figure~\ref{figure:cc-distance} shows a positive correlation between the CC of the input programs and the Levenshtein distance, as the correlation coefficient is 0.4886.
It indicates that the LLM makes more extensive changes to the program with larger CC trying to make it less complex, whereas the LLM makes only small changes if the given program is already simple.
Similarly, Figure~\ref{figure:chars-distance} shows a strong positive correlation between the number of characters of the input programs and the Levenshtein distance, as the correlation coefficient is 0.7442.
It indicates that the LLM makes longer changes to the longer programs.

\subsubsection{Decrease in LOC, Chars, and Tokens}

\begin{table}[h]
\begin{center}
\begin{minipage}{\linewidth}
    \caption{Average values of LOC, Chars, and Tokens of the programs. The values in parentheses indicate the standard deviation.}
    \label{table:lengths}
    \begin{tabular*}{\textwidth}{@{\extracolsep{\fill}}lccc@{\extracolsep{\fill}}}
        \toprule
        & LOC & Chars & Tokens \\
        \midrule
        Input & 14.82 ($\pm$ 17.29) & 331.35 ($\pm$ 462.24) & 106.15 ($\pm$ 116.70) \\
        Generated & 10.99 ($\pm$ 10.76) & 263.17 ($\pm$ 333.94) & 85.18 ($\pm$ 95.68) \\
        \bottomrule
    \end{tabular*}
\end{minipage}
\end{center}
\end{table}

As shown in Table~\ref{table:lengths}, all of the metrics of LOC, Char-based length, and Token-based length are reduced in the generated programs compared with the input programs.
Although the metrics do not directly indicate the readability and maintainability of the programs (i.e., longer programs may have descriptive variable names and comments), it shows that the generated programs are more concise while keeping the semantic correctness and reducing the complexity.

\subsubsection{Levenshtein Distance and Similarity}

\begin{table}[h]
\begin{center}
\begin{minipage}{0.8\linewidth}
    \caption{Character-based Levenshtein distance and similarity, compared with the input programs.}
    \label{table:distance}
    \begin{tabular*}{\textwidth}{@{\extracolsep{\fill}}lcc@{\extracolsep{\fill}}}
        \toprule
        & Mean (Std) & Min $\sim$ Max \\
        \midrule
        Distance & 95.59 ($\pm$ 162.89) & 1 $\sim$ 3035 \\
        Similarity & 67.54\% ($\pm$ 16.95\%) & 10.76\% $\sim$ 99.46\% \\
        \bottomrule
    \end{tabular*}
\end{minipage}
\end{center}
\end{table}

In addition, as shown in Table~\ref{table:distance}, we identify that the generated programs contain at least one or more edits, as the minimum Levenshtein distance is 1 and the maximum similarity is 99.46\%.
This further indicates that there is no \textit{cheating case} where the correct program is just returned without any edits to pass the semantic validation.
However, on the other hand, this suggests that modifications can be forcibly made, even if the input program is already readable.

\subsection{Qualitative}
\label{sec:results:qualitative}

\subsubsection{Main Results}

For a qualitative evaluation, we manually inspect the generated programs randomly.
Overall, the suggested programs contain helpful modifications to make the input programs less complex, as well as the improvement of readability and maintainability.
For the readability improvement, we observe the cases of variable renaming for more descriptive variables and code formatting of adding or removing white spaces and white lines appropriately.

\begin{figure}
    \centering
    \begin{subfigure}[b]{0.54\linewidth}
        \centering
        \includegraphics[width=\linewidth]{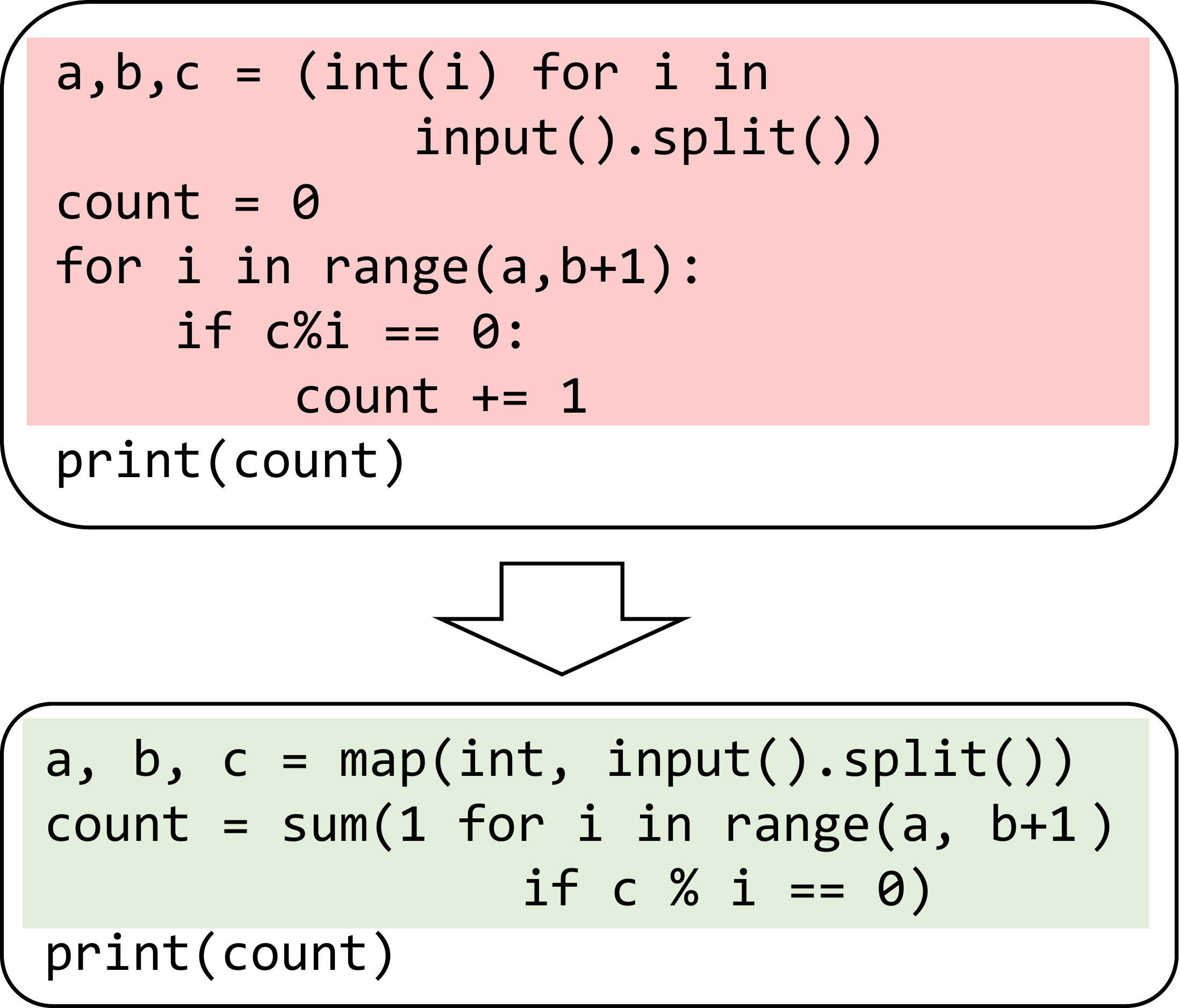}
        \caption{Improved example.}
        \label{figure:improved-refactoring}
    \end{subfigure}
    \hfill
    \begin{subfigure}[b]{0.44\linewidth}
        \centering
        \includegraphics[width=\linewidth]{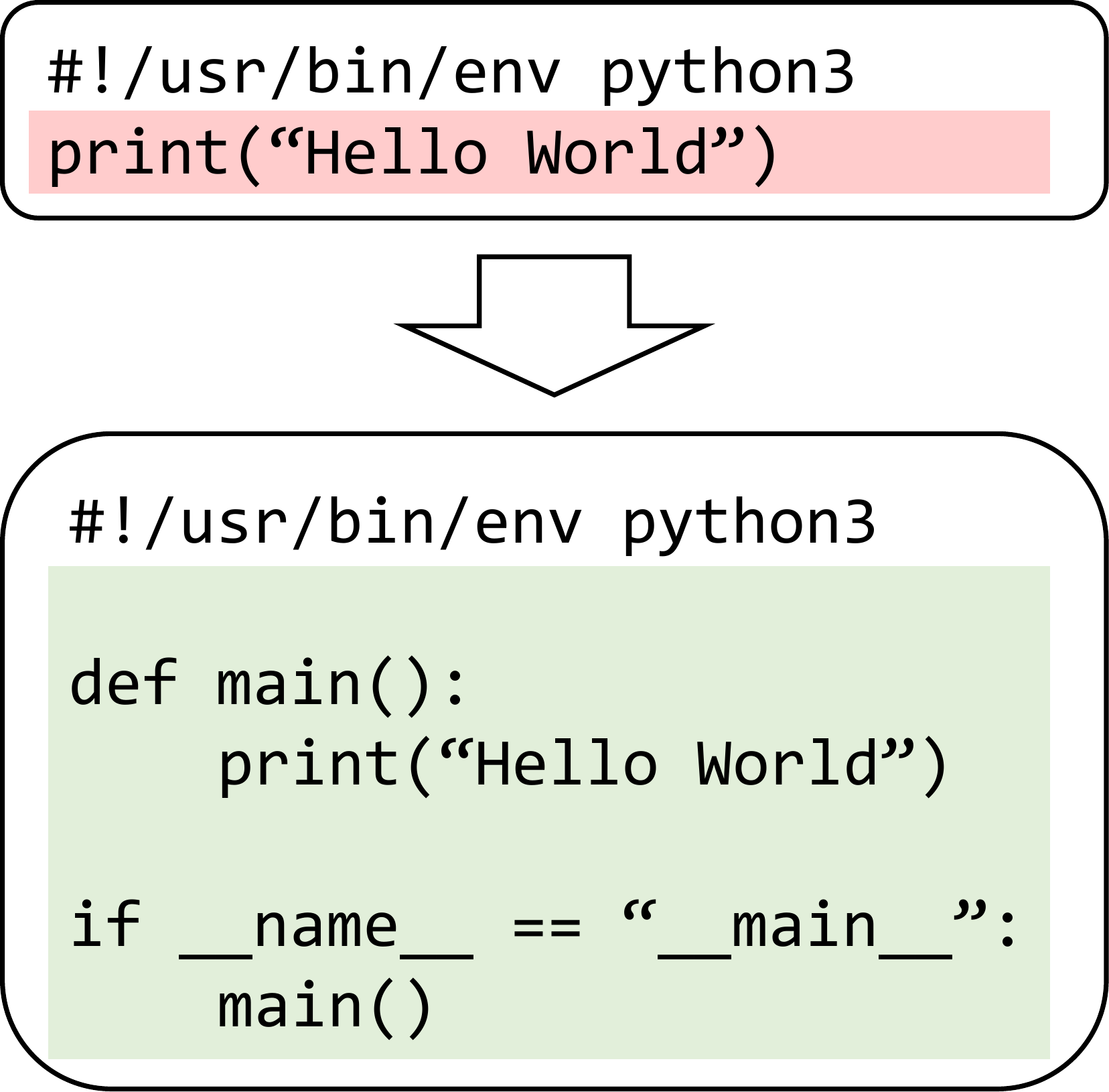}
        \caption{Worsened example.}
        \label{figure:worsened-refactoring}
    \end{subfigure}
    \caption{Examples of refactoring by the LLM.}
    \label{figure:refactoring-examples}
\end{figure}

Figure~\ref{figure:refactoring-examples} shows the representatives of improved and worsened refactorings generated by the LLM.
Figure~\ref{figure:improved-refactoring} is the improved example, changing from for and if statements to an effective generator.
Figure~\ref{figure:worsened-refactoring} is the worsened example, defining an excessive function that worsens the complexity.

\subsubsection{Code Formatting}

To support the observation of the capability in code formatting, we perform additional analysis to calculate the character-based Levenshtein distance between the programs and their formatted programs.
We employ the \texttt{yapf}\footnote{\url{https://github.com/google/yapf}.} library to format the Python programs.
While the input programs have a 21.96 ($\pm$ 50.83) distance from the formatted programs on average, the generated programs have only a 7.95 ($\pm$ 32.78) distance.
The significantly smaller distance in the generated programs indicates that the generated programs are well-formatted and less required to be formatted.

\subsubsection{Comment Deletion}

\begin{table}[h]
\begin{center}
\begin{minipage}{\linewidth}
    \caption{The average number of comments in the input and generated programs. Incl. and excl. indicate whether to include inline comments. The comment ratio is calculated by dividing the number of comment lines excluding inline comments by the number of lines of code including white lines.}
    \label{table:comments}
    \begin{tabular*}{\textwidth}{@{\extracolsep{\fill}}lccc@{\extracolsep{\fill}}}
        \toprule
        & Comments (incl.) & Comments (excl.) & Comment Ratio \\
        \midrule
        Input & 0.60 ($\pm$ 2.59) & 0.53 ($\pm$ 2.44) & 2.39\% ($\pm$ 7.92\%) \\
        Generated & 0.16 ($\pm$ 0.75) & 0.14 ($\pm$ 0.70) & 0.89\% ($\pm$ 4.42\%) \\
        \bottomrule
    \end{tabular*}
\end{minipage}
\end{center}
\end{table}

Although we observe many cases improving the code readability, we also observe that the LLM deletes many comments that can help understand the program.
We count the number of comments using the standard \texttt{tokenize}\footnote{\url{https://docs.python.org/3/library/tokenize.html}} library to support this observation.
As shown in Table~\ref{table:comments}, we verify that the average number of comment lines decreased from 0.60 to 0.16, which reduced the comment ratio from 2.39\% to 0.89\%.

\subsubsection{Comment Translation}

\begin{table}[h]
\begin{center}
\begin{minipage}{.8\linewidth}
    \caption{The proportion of natural languages used in code.}
    \label{table:languages}
    \begin{tabular*}{\textwidth}{@{\extracolsep{\fill}}clrlr@{\extracolsep{\fill}}}
        \toprule
        & \multicolumn{2}{c}{Input} & \multicolumn{2}{c}{Generated} \\
        \midrule
        1 & English & 95.11\% & English & 97.16\% \\
        2 & Japanese & 3.98\% & Japanese & 1.58\% \\
        3 & Unknown & 0.91\% & Unknown & 1.26\% \\
        \bottomrule
    \end{tabular*}
\end{minipage}
\end{center}
\end{table}

Similar to deleting comments, we also observe a few cases of non-English comments (e.g., Japanese) being translated into English without a demand.
To support this observation, we count the languages used in the code by asking GPT-3.5 to determine the natural language.
We use the engine \texttt{gpt-3.5-turbo-0301} with the parameters of $temperature = 0$ to be deterministic and $max\_tokens = 10$ to be answered concisely. We use the system instruction of \textit{``What natural language is used in this code? Select from [English, Japanese, Korean, Chinese, Other, Unknown, None].''} to be easier to aggregate.
We compare the changes in the language proportion since it is difficult to quantitatively detect what language is translated into what language for each comment, as the whole program is refactored by the LLM.
The decrease in the proportion of Japanese and the increase in English in Table~\ref{table:languages} suggests that the Japanese comments are translated into English without demand.
Note that the reason only English or Japanese comments are used is that the AOJ is a Japanese web service that supports only English and Japanese.

However, the decrease in the number of comments (i.e., natural language explanations) and the preferred use of English might be affected by the system instruction we used.
Firstly, the instruction \textit{``Do not explain anything in natural language.''} might suppress the natural language explanations in the comments, although we intended to suppress explanations outside the code block (program).
Secondly, the LLM might decide that the comments should be rewritten in English so that the users using English can understand them since the instruction is written in English.

\section{Limitations}
\label{sec:limitations}

As described in Section~\ref{sec:results:quantitative}, the LLM makes at least some modifications to the given program.
It can worsen an already straightforward, readable, or concise program that is not required to be refactored.
To avoid this, applying some techniques to make modifications only if the program requires refactoring can be beneficial.
For instance, detecting whether the program should be refactored before the refactoring request, or providing some instructions to control the LLM to make modifications only when there is a particular need for refactoring, can be considered.
Furthermore, since code refactoring is usually performed iteratively, suggesting the fully refactored program with extensive modifications at once can be overwhelming for users.
Iterative suggestions containing a small fraction of the modifications would be more educational.

In this work, we employ GPT-3.5 as the representative LLM from the publicly available LLMs to demonstrate the effectiveness of our proposed approach.
Many LLMs capable of solving programming problems have been proposed recently~\cite{touvron2023llama, openai2023gpt4, zheng2023codegeex, li2023starcoder, luo2023wizardcoder}, including those that may perform even better than GPT-3.5.
However, our results, which show the refactoring ability of an LLM that can be further improved by selecting better few-shot examples, can be a foundation for further advancements in applying LLMs for code refactoring tasks.

\section{Conclusion}
\label{sec:conclusion}

This paper presents a methodology for using a large language model (LLM), GPT-3.5, to suggest less complex versions of user-written Python programs.
The proposed approach leverages few-shot prompting with carefully selected examples to encourage users to learn how to write better programs.

The quantitative evaluation demonstrates that the LLM can generate correct and less complex programs for a majority (95.68\%) of the input programs by generating 10 candidates for each.
The average cyclomatic complexity is reduced by 17.35\%, and the average number of lines of code is decreased by 25.84\%, indicating the effectiveness of the LLM in code refactoring.
Furthermore, we observe several associations in metrics, suggesting the capabilities and limitations of the LLM.

The qualitative evaluation shows that the LLM can improve code readability through variable renaming and formatting.
However, we also observe that the LLM made unnecessary modifications to already straightforward and readable programs, as well as deletion and translation of comments.

Overall, this work demonstrates the potential of LLMs in supporting code refactoring and improving program complexity. Future research includes exploring using different LLMs and further improvements on making code refactoring examples to enhance the pass rates and effectiveness of the suggested refactorings.

\bibliographystyle{ieeetr}
\bibliography{main}

\end{document}